\documentclass[preprint,nofootinbib]{revtex4}%
\usepackage{amssymb}
\usepackage{amsfonts}
\usepackage{amsmath}
\usepackage{graphicx}
\usepackage[usenames]{color}%
\providecommand{\U}[1]{\protect\rule{.1in}{.1in}}
\definecolor{blue}{rgb}{0,0,1}

\definecolor{red}{rgb}{1,0,0}

\begin{document}
\title{Asymptotically locally AdS and flat black holes \\in Horndeski theory}
\author{Andres Anabalon$^{1,2}$, Adolfo Cisterna$^{3}$ and Julio Oliva$^{4}$}
\affiliation{$^{1}${\small Departamento de Ciencias, Facultad de Artes Liberales y Facultad
de Ingenier\'{\i}a y Ciencias, Universidad Adolfo Ib\'{a}\~{n}ez, Vi\~{n}a del
Mar, Chile.}}
\affiliation{$^{2}${\small Universit\'{e} de Lyon, Laboratoire de Physique, UMR 5672, CNRS,
\'{E}cole Normale Sup\'{e}rieure de Lyon, 46 all\'{e} d'Italie, F-69364 Lyon
Cedex 07, France.}}
\affiliation{$^{3}${\small Instituto de F\'{\i}sica, Pontificia Universidad de Cat\'{o}lica
de Valpara\'{\i}so, Av. Universidad 330, Curauma, Valpara\'{\i}so, Chile}}
\affiliation{$^{4}${\small Instituto de Ciencias F\'{\i}sicas y Matem\'{a}ticas,
Universidad Austral de Chile, Valdivia, Chile}\footnote{andres.anabalon@uai.cl
- adolfo.cisterna.r@mail.pucv.cl - julio.oliva@uach.cl}}

\begin{abstract}
In this paper we construct asymptotically locally AdS and flat black holes in
the presence of a scalar field whose kinetic term is constructed out from a
linear combination of the metric and the Einstein tensor. The field equations
as well as the energy-momentum tensor are second order in the metric and the
field, therefore the theory belongs to the ones defined by Horndeski. We show
that in the presence of a cosmological term in the action, it is possible to
have a real scalar field in the region outside the event horizon. The
solutions are characterized by a single integration constant, the scalar field
vanishes at the horizon and it contributes to the effective cosmological
constant at infinity. We extend these results to the topological case. The
solution is disconnected from the maximally symmetric AdS background, however,
within this family there exits a gravitational soliton which is everywhere
regular. This soliton is therefore used as a background to define a finite
Euclidean action and to obtain the thermodynamics of the black holes. For a
certain region in the space of parameters, the thermodynamic analysis reveals
a critical temperature at which a Hawking-Page phase transition between the
black hole and the soliton occurs. We extend the solution to arbitrary
dimensions grater than four and show that the presence of a cosmological term
in the action allows to consider the case in which the standard kinetic term
for the scalar it's not present. In such scenario, the solution reduces to an
asymptotically flat black hole.

\end{abstract}
\maketitle

\section{Introduction}

In many scenarios, the gravitational dynamics is successfully described by
Einstein's theory of General Relativity (GR), which has a large observational
support \cite{Damour}. However there is also a strong empirical motivation,
mainly associated with astronomical observations, that alternative theories of
gravity could be useful in order to describe phenomena such as dark matter and
dark energy, or the process of inflation suffered by the early Universe. Many
of these theories are based in the use scalar fields. The inclusion of scalar
fields in physics has a vital role. Scalar fields for example appear in the
Brans-Dicke theory as an attempt to incorporate Mach's principle in a
gravitational theory \cite{BD}, in inflationary theories, and as a possible
candidate of dark matter. On the other hand in particle physics, the Higgs
boson is a fundamental piece of the standard model, and its recent discovery
gives a firm ground to the realization of scalar fields in nature. Moreover,
scalar fields are also ubiquitous in string theory and Kaluza-Klein
compactifications of higher dimensional theories.

The study of scalar-tensor theories have recently attracted considerable
attention due to the development of Galileon theories and their applications
\cite{NRT}. In particular, these ideas led the re-discovery of the most
general scalar-tensor theory which has second order field equations and second
order energy-momentum tensor; a problem that was solved by Horndeski
\cite{Horndeski} in the early seventies. In a curved four-dimensional
background, the most general Lagrangian that can be constructed with these
properties is given by:%
\begin{equation}
L=\beta_{1}\delta_{efhi}^{abdc}R_{ab}^{ef}R_{cd}^{hi}+\beta_{2}\delta
_{def}^{acd}\nabla_{a}\phi\nabla^{d}\phi R_{cd}^{ef}+\beta_{3}\delta_{cd}%
^{ab}R_{ab}^{cd}+\Xi+C\epsilon^{abdc}R^{p}{}_{qab}R^{q}{}_{pcd}\label{horn4}%
\end{equation}
where $C$ is a constant, $\beta_{i}$ are arbitrary functions of the scalar
$\phi$ and $\Xi$ is an arbitrary function of the scalar field and its squared
gradient, i.e. $\Xi=\Xi(\nabla_{a}\phi\nabla^{a}\phi,\phi)$. The first term is
a nonminimal coupling between the scalar and the four-dimensional Gauss-Bonnet
density, the second includes a nonminimal coupling between the standard
kinetic term and the Einstein tensor, the third one is a nonminimal coupling
between the field and the Ricci scalar, while the fourth term since it is
defined by an arbitrary function it might include an additive term which may
act as a cosmological term in the action. The last term is just the
Pointryagin term which is a boundary term and does not contribute to the field
equations\footnote{An interesting application of this Lagrangian in the
comological setup has been given in \cite{Charmousis:2011bf} where it can be seen that a
particular case of Horndeski action provides a novel self tunning mechanism.}.
From here we see that it is possible to have scalar field Lagrangians whose
kinetic term has non-minimal derivative couplings with the curvature.

Let us consider kinetic terms $H$ which are quadratic in the derivatives of
the field in arbitrary dimension $D$. Requiring second order energy-momentum
tensor as well as linear and second order equation for the field, restricts
$H$ to be a linear combination of the following basic terms:%
\begin{equation}
H^{\left(  n\right)  }=E_{\mu\nu}^{(n)}\nabla^{\mu}\phi\nabla^{\nu}%
\phi\ ,\label{familyn}%
\end{equation}
where $E_{\mu\nu}^{(n)}$ is the $n$-$th$ order Lovelock tensors\footnote{The
most general symmetric tensors which are divergency-free and contain up to
second order derivatives of the metric.}
\begin{equation}
E^{(n)\ }{}_{\mu}^{\nu}=\delta_{\mu\beta_{1}...\beta_{2n}}^{\nu\alpha
_{1}...\alpha_{2n}}R_{\ \ \alpha_{1}\alpha_{2}}^{\beta_{1}\beta_{2}%
}...R_{\ \ \alpha_{2n-1}\alpha_{2n}}^{\beta_{2n-1}\beta_{2n}}\
\end{equation}
The standard kinetic term is recovered with $n=0$. Since $E_{\mu\nu}^{\left(
1\right)  }$ is proportional to the Einstein tensor, the first non-standard
term in (\ref{familyn}) already includes a non-minimal kinetic coupling of the
scalar and the curvature.

In this work we will consider the following action principle:%

\begin{equation}
I[g_{\mu\nu},\phi]=\int\sqrt{-g}d^{D}x\ \left[  \kappa\left(  R-2\Lambda
\right)  -\frac{1}{2}\left(  \alpha g_{\mu\nu}-\eta G_{\mu\nu}\right)
\nabla^{\mu}\phi\nabla^{\nu}\phi\right]  \label{action}%
\end{equation}
where we have taken the Einstein-Hilbert action with a cosmological term in
the gravity sector, while the matter sector is given by a real scalar field
with a non-minimal kinetic coupling (here $\kappa:=\frac{1}{16\pi G}$). The
possible values of the dimensionfull parameters $\alpha$ and $\eta$ will be
determined below according with the positivity of the energy density of the
matter field.

Numerical solutions in the case when an electromagnetic field is present were
found in \cite{NumericalSol}, where phase transitions to charged black holes
with complex anisotropic scalar hair were explored. The first exact black hole
solution to this system, in the case of a vanishing cosmological term
$\Lambda$, was found by Rinaldi in \cite{Rinaldi}. There, the scalar field
becomes imaginary in the domain of outer communication, and the weak energy
condition is violated outside the horizon.

\bigskip

Here we extend the results of \cite{Rinaldi} and show that the inclusion of a
cosmological term in the action makes possible finding a black hole with a
real scalar field outside the horizon. By a suitable regularization of the
action, we explore the thermal properties of the spherically symmetric black
holes. We show that for a region in the space of parameters, there is a phase
transition as in General Relativity with a negative cosmological constant  We
also extend the solution to the topological case in arbitrary dimension
$D\geq4$ and show that the cosmological term allows to obtain a non-trivial
solution when $\alpha=0$. In this case we obtain an asymptotically flat black hole.

\bigskip

The field equations for the metric and the scalar field are:%

\begin{equation}
G_{\mu\nu}+\Lambda g_{\mu\nu}=\frac{\alpha}{2\kappa}T_{\mu\nu}^{(1)}%
+\frac{\eta}{2\kappa}T_{\mu\nu}^{\left(  2\right)  } \label{eqmetric}%
\end{equation}

\begin{equation}
\nabla_{\mu}\left[  \left(  \alpha g^{\mu\nu}-\eta G^{\mu\nu}\right)
\nabla_{\nu}\phi\right]  =0\label{eqphi}%
\end{equation}
where\footnote{We use a normalized symmetrization $A_{(\mu\nu)}:=\frac{1}%
{2}\left(  A_{\mu\nu}+A_{\nu\mu}\right)  $.}:%
\begin{align*}
T_{\mu\nu}^{\left(  1\right)  } &  =\nabla_{\mu}\phi\nabla_{\nu}\phi-\frac
{1}{2}g_{\mu\nu}\nabla_{\lambda}\phi\nabla^{\lambda}\phi\\
T_{\mu\nu}^{\left(  2\right)  } &  =\frac{1}{2}\nabla_{\mu}\phi\nabla_{\nu
}\phi R-2\nabla_{\lambda}\phi\nabla_{(\mu}\phi R_{\nu)}^{\lambda}%
-\nabla^{\lambda}\phi\nabla^{\rho}\phi R_{\mu\lambda\nu\rho}\\
&  -(\nabla_{\mu}\nabla^{\lambda}\phi)(\nabla_{\nu}\nabla_{\lambda}%
\phi)+(\nabla_{\mu}\nabla_{\nu}\phi)\square\phi+\frac{1}{2}G_{\mu\nu}%
(\nabla\phi)^{2}\\
&  -g_{\mu\nu}\left[  -\frac{1}{2}(\nabla^{\lambda}\nabla^{\rho}\phi
)(\nabla_{\lambda}\nabla_{\rho}\phi)+\frac{1}{2}(\square\phi)^{2}%
-\nabla_{\lambda}\phi\nabla_{\rho}\phi R^{\lambda\rho}\right]
\end{align*}
We will consider a gauge fixed version of the most general, cohomogeneity one,
static spacetime (which is not a direct product):%
\begin{equation}
ds^{2}=-F(r)dt^{2}+G(r)dr^{2}+r^{2}d\Sigma_{K}^{2}\label{metric}%
\end{equation}
where $d\Sigma_{K}$ is the line element a closed, $(D-2)$-dimensional
Euclidean space of a constant curvature $K=0,\pm1$. For $K=1$, the space
$\Sigma_{K}$ is locally a sphere, for $K=0$ it is locally flat and for $K=-1$
it locally reduces to a hyperbolic space. Hereafter we will consider a static
and isotropic scalar field, i.e. $\phi=\phi\left(  r\right)  $.

The outline of the paper is as follows: in section 2 the four-dimensional
solution is given for arbitrary $K$. In order to constraint the couplings, the
weak energy condition is imposed on the scalar field. In section 3, the
geometry of the spherically symmetric solution is described in detail as well
it's thermal properties. We compute the temperature, the mass and the entropy
following the Hawking-Page approach, and show that the first law is satisfied.
We prove that further restrictions on the integration constant come from
requiring the black hole to have positive entropy, and show that for a certain
region in the space of parameters there is a Hawking-Page phase transition. In
section 4, the solution in arbitrary dimension $D$ is given. Finally in
section 5, the solution in the special case $\alpha=0$ is analyzed, which
gives rise to an asymptotically locally flat black hole. In this paper we use
the signature $(-,+,+,+)$, Greek and Latin indices stand for indices in the
coordinate basis.

\bigskip

\textit{Note added: while the first version of this work was being finished
the article \cite{BC} appeared on ArXiv which has some overlap with our four
dimensional solution.}

\section{Four-dimensional solution}

For the static ansatz considered here, one has that the equation for the field
(\ref{eqphi}) admits a first integral, which implies the relation:%
\begin{equation}
r\frac{F^{\prime}(r)}{F(r)}=\left[  K+\frac{\alpha}{\eta}r^{2}-\frac{C_{0}%
}{\eta}\frac{G(r)}{\psi(r)\sqrt{F(r)G(r)}}\right]  G(r)-1\ ,
\label{eqfieldfixed}%
\end{equation}
where $C_{0}$ is an integration constant, $\psi(r):=\phi^{\prime}(r)$, and
$(^{\prime})$ stands for derivation with respect to $r$. As in reference
\cite{Rinaldi}, in order to obtain an exact solution we (arbitrarily) choose
$C_{0}=0$. Therefore, equation (\ref{eqfieldfixed}) in addition to the $tt$
and $rr$ components of (\ref{eqmetric}), provides a consistent system which
for $K=\pm1$ and $\eta\Lambda\neq\alpha$, has the following solution%

\begin{equation}
F(r)=\frac{r^{2}}{l^{2}}+\frac{K}{\alpha}\sqrt{\alpha\eta K}\left(
\frac{\alpha+\Lambda\eta}{\alpha-\Lambda\eta}\right)  ^{2}\frac{\arctan\left(
\frac{\sqrt{\alpha\eta K}}{\eta K}r\right)  }{r}-\frac{\mu}{r}+\frac
{3\alpha+\Lambda\eta}{\alpha-\eta\Lambda}K\ , \label{Fsol}%
\end{equation}

\begin{equation}
G(r)=\frac{\alpha^{2}(\left(  \alpha-\eta\Lambda\right)  r^{2}+2\eta K)^{2}%
}{(\alpha-\eta\Lambda)^{2}(\alpha r^{2}+\eta K)^{2}F(r)}\ , \label{Gsol}%
\end{equation}

\begin{equation}
\psi^{2}(r)=-\frac{2r^{2}\kappa\alpha^{2}(\alpha+\eta\Lambda)(\left(
\alpha-\eta\Lambda\right)  r^{2}+2\eta K)^{2}}{\eta(\alpha-\eta\Lambda
)^{2}(\alpha r^{2}+\eta K)^{3}F(r)}\ . \label{psisol}%
\end{equation}
\ where $\mu$ is an integration constant. Here we have defined the effective
(A)dS radius $l$ by
\begin{equation}
l^{-2}:=\frac{\alpha}{3\eta}\ . \label{defl}%
\end{equation}
In the case of a locally flat transverse section ($K=0$) the solution reduces
to topological Schwarzschild solution with locally flat horizons
\cite{Lemos:1994xp}:%

\begin{equation}
F(r)=\frac{r^{2}}{l^{2}}-\frac{\mu}{r}=\frac{1}{G\left(  r\right)  }%
\end{equation}

\begin{equation}
\psi(r)^{2}=-\frac{2\kappa(\alpha+\eta\Lambda)}{\alpha\eta}\frac{1}{F\left(
r\right)  }%
\end{equation}
and can be obtained from the one with arbitrary $K$ by taking the formal limit
$K\rightarrow0$. Note that in both cases, the solution obtained by replacing
$\mu\rightarrow-\mu$ is the same than the former if we replace as well
$r\rightarrow-r$.

As mentioned above, from equation (\ref{psisol}) one can see explicitly that
the sign of $\psi^{2}\left(  r\right)  $ is determined by the sign of the
combination $\frac{(\alpha+\eta\Lambda)}{\alpha\eta}$, and therefore
regardless the sign of $\eta/\alpha$, the inclusion of a cosmological term in
the action allows for the existence of asymptotically (A)dS black holes with a
real scalar field in the outer domain of communication.

\bigskip

It can be seen that this solution is asymptotically locally dS or AdS for
$\alpha/\eta<0$ or $\alpha/\eta>0$, respectively, since when $r\rightarrow
\infty$ the components of the Riemann tensor go to%
\begin{equation}
R_{\ \ \lambda\sigma}^{\mu\nu}\underset{r\rightarrow\infty}{\sim}-\frac
{\alpha}{3\eta}\delta_{\lambda\sigma}^{\mu\nu}=:-\frac{1}{l^{2}}%
\delta_{\lambda\sigma}^{\mu\nu}\ ,
\end{equation}
justifying our previous definition of the effective (A)dS radius (\ref{defl}).
The asymptotic expansion ($r\rightarrow\infty$) of the metric functions is%
\begin{align}
&  -g_{tt}\underset{r\rightarrow\infty}{=}\frac{r^{2}}{l^{2}}+\frac
{3\alpha+\eta\Lambda}{\alpha-\eta\Lambda}K+\left(  \frac{K}{2\alpha}%
\sqrt{\alpha\eta K}\left(  \frac{\alpha+\eta\Lambda}{\alpha-\eta\Lambda
}\right)  ^{2}\pi\sigma-\mu\right)  \frac{1}{r}+O\left(  r^{-2}\right)  \ ,\\
&  g^{rr}\underset{r\rightarrow\infty}{=}\frac{r^{2}}{l^{2}}+\frac
{7\alpha+\eta\Lambda}{3(\alpha-\eta\Lambda)}K+\left(  \frac{K}{2\alpha}%
\sqrt{\alpha\eta K}\left(  \frac{\alpha+\eta\Lambda}{\alpha-\eta\Lambda
}\right)  ^{2}\pi\sigma-\mu\right)  \frac{1}{r}+O\left(  r^{-2}\right)  \ ,
\end{align}
where $\sigma$ is the sign of the combination $\eta K$. For $\mu\neq0$, there
is a curvature singularity at $r=0$ since for example the Ricci scalar
diverges as%
\begin{equation}
R\underset{r\rightarrow0}{=}\frac{3\left(  \alpha+\Lambda\eta\right)  \left(
\alpha-\Lambda\eta\right)  ^{2}}{4\eta K\alpha^{2}}\frac{\mu}{r}+O\left(
1\right)  \ .
\end{equation}

\bigskip

If the combination $(\alpha+\eta\Lambda)$ vanishes, one can see that the
scalar field reduces to a constant and the metric reduces to the topological
Schwarzschild-AdS solution of General Relativity.

For $(\alpha+\eta\Lambda)\neq0$, the metric is disconnected from the maximally
symmetric AdS vacua, i.e. that in this case it is not possible to set $\mu$ to
some value such that the metric reduces to AdS. Nevertheless for $\mu=0$ it is
possible to show that all the components of the Riemann tensor $R_{\ \ cd}%
^{ab}$ are finite when $r\rightarrow0$. All the algebraic curvature invariants
$\mathcal{I}$, i.e. those constructed out from contractions of the Riemann
tensor without involving covariant derivatives, can be written as linear
combinations of products of the components of the Riemann tensor with the
index structure $R_{\ \ cd}^{ab}$ without involving metric factor. Therefore,
the finiteness of the components of the Riemann tensor with this index
structure ensures the finiteness of all the algebraic curvature invariants.
For $\mu=0$ the scalar field is finite at the origin, therefore invariants of
the form $\phi^{p}\times\mathcal{I}$ will be finite for arbitrary $p$. This
has important consequences, because it suggests considering the spacetime with
$\mu=0$ as a background to define a regularized Euclidean action. As described
below, the solution with $\mu=0$ actually defines a gravitational soliton.

If $\rho(r)$ is the energy density, then the total energy $\mathcal{E}$\ is
given by:%
\begin{equation}
\mathcal{E}=V\left(  \Sigma\right)  \int r^{2}\sqrt{G\left(  r\right)  }%
dr\rho\left(  r\right)  \ ,
\end{equation}
where $V\left(  \Sigma\right)  $ stands for the volume of $\Sigma$. Therefore:%
\begin{equation}
T_{00}=\rho\left(  r\right)  :=F\left(  r\right)  ^{-1}T_{tt}\
\end{equation}
where $T_{00}=e_{0}^{\mu}e_{0}^{\nu}T_{tt}$ and $e_{\mu}^{a}$ is a local
frame. Now, the $tt$ component of the energy momentum tensor reads:%
\begin{equation}
T_{tt}=-\frac{\kappa^{2}\left(  \alpha+\Lambda\eta\right)  }{\eta}F\left(
r\right)  \left[  \frac{4K\eta^{2}\left(  \alpha-\Lambda\eta\right)
^{2}\left(  \alpha r^{2}+\eta K\right)  }{\alpha^{2}\left(  \left(
\alpha-\Lambda\eta\right)  r^{2}+2\eta K\right)  ^{3}}F\left(  r\right)
+1\right]  \ .
\end{equation}
As expected for a matter action that is quadratic in the derivatives of the
field, the positivity of the energy density is in close relation with the
reality of the scalar itself, which is determined by the sign of the
combination $\frac{(\alpha+\eta\Lambda)}{\alpha\eta}$ as can be seen from
equation (\ref{psisol}). In the following section we analyze in detail the
case $K=1$ and compute the thermal properties of these black holes.

\bigskip

\section{Spherically symmetric case}

In the spherically symmetric case, the metric functions and the square of the
derivative of the scalar field take the form:%

\begin{equation}
F(r)=\frac{r^{2}}{l^{2}}+\frac{1}{\alpha}\sqrt{\alpha\eta}\left(  \frac
{\alpha+\Lambda\eta}{\alpha-\Lambda\eta}\right)  ^{2}\frac{\arctan\left(
\frac{\sqrt{\alpha\eta}}{\eta}r\right)  }{r}-\frac{\mu}{r}+\frac
{3\alpha+\Lambda\eta}{\alpha-\Lambda\eta}\ , \label{Fsph}%
\end{equation}%
\begin{equation}
G(r)=\frac{\alpha^{2}(\left(  \alpha-\eta\Lambda\right)  r^{2}+2\eta)^{2}%
}{(\alpha-\eta\Lambda)^{2}(\alpha r^{2}+\eta)^{2}F(r)}\ , \label{Gsph}%
\end{equation}

\begin{equation}
\psi^{2}(r)=-\frac{2r^{2}\kappa\alpha^{2}(\alpha+\eta\Lambda)(\left(
\alpha-\eta\Lambda\right)  r^{2}+2\eta)^{2}}{\eta(\alpha-\eta\Lambda
)^{2}(\alpha r^{2}+\eta)^{3}F(r)}\ .\label{psisph}%
\end{equation}
Since the lapse function has a term including the combination $\sqrt
{\alpha\eta}$, we can see that $\alpha$ and $\eta$ need to have the same sign;
$l^{-2}:=\frac{\alpha}{3\eta}$ will be positive definite and therefore the
spacetime is asymptotically locally AdS. Without loosing generality, we
consider $\alpha$ and $\eta$ positive, since the solution with both $\alpha$
and $\eta$ negative is equivalent to the former by changing $\mu
\rightarrow-\mu$.

The reality of the field in the asymptotic region implies $(\alpha+\eta
\Lambda)<0$, which in our case imposes $\Lambda<-\frac{\alpha}{\eta}<0$. Under
these conditions one can see that the solution describes a black hole with a
non-degenerate horizon for $\mu>0$, which is located at $r=r_{+}$. Since the
horizon is non-degenerate, $F\left(  r_{+}\right)  =0$ while $F^{\prime
}\left(  r_{+}\right)  \neq0$, therefore close to the horizon, we have:
\[
\psi\left(  r\right)  \underset{r\rightarrow r_{+}}{\sim}\frac{\zeta}%
{2\sqrt{r-r_{+}}}\Rightarrow\phi\left(  r\right)  \underset{r\rightarrow
r_{+}}{\sim}\zeta\sqrt{r-r_{+}}\text{\ ,}%
\]
for some constant $\zeta$, which implies that the scalar field vanishes at the
horizon and it is non-analytic there.

\bigskip

If $(\alpha+\eta\Lambda)\neq0$, it is important to realize that it is not
possible to switch off the scalar field and therefore our solution is not
continuously connected with a globally maximally symmetric background.
Nevertheless it is possible to show, as it occurs in the $\Lambda=0$ case
\cite{Rinaldi}, that within this family, the only regular spacetime is
obtained by setting $\mu=0$. In this case the metric describes an
asymptotically AdS gravitational soliton. Near $r=0$, after a proper rescaling
of the time coordinate, the soliton metric reduces to:%
\begin{equation}
ds_{soliton}^{2}\underset{r\rightarrow0}{\sim}-\left(  1-\frac{\Lambda}%
{3}r^{2}+O(r^{4})\right)  dt^{2}+\left(  1-\frac{3\alpha+2\Lambda\eta}{3\eta
}r^{2}+O(r^{4})\right)  dr^{2}+r^{2}d\Omega^{2}\;\text{,}%
\end{equation}
therefore one can see explicitly that it has a regular origin.

The thermal version of such spacetime can be considered as the background
metric to obtain a regularized Euclidean action along the lines of
Hawking-Page \cite{HP}, and to obtain therefore the thermodynamics of the
black holes. This will be done in the next section.

\subsection{Thermodynamics of the spherically symmetric black holes}

The regularized Euclidean action is defined by:%
\begin{equation}
I_{reg}=I_{E}\left[  g_{\mu\nu},\phi\right]  -I_{E}\left[  g_{\mu\nu}^{\left(
0\right)  },\phi^{\left(  0\right)  }\right]  \ ,\label{regul action}%
\end{equation}
where $I_{E}$ and $g_{\mu\nu}^{\left(  0\right)  }$ and $\phi^{\left(
0\right)  }$ are the metric and the scalar field respectively for the
gravitational soliton, which are obtained by setting $\mu=0$ in the black hole
solution defined by the functions (\ref{Fsph}) to (\ref{psisph}). Removing the
conical singularity at the horizon of the Euclidean black hole solution,
requires the period of the Euclidean time to be fixed as%
\begin{equation}
\beta=\left.  4\pi\sqrt{\frac{G^{\prime}}{F^{\prime}}}\right\vert _{r=r_{+}%
}=\frac{4\pi\eta\left(  \alpha-\eta\Lambda\right)  r_{+}}{\alpha\left(
2\eta+\left(  \alpha-\Lambda\eta\right)  r_{+}^{2}\right)  }\ .
\end{equation}
In order for equation (\ref{regul action}) to define a regularized action we
need to consider a thermal gravitational soliton. If we denote the period of
the Euclidean time for the soliton by $\beta_{0}$, we need to impose the
redshifted temperatures to match:
\begin{equation}
\beta^{2}F\left(  r=r_{c},\mu\right)  =\beta_{0}^{2}F\left(  r=r_{c}%
,\mu=0\right)  \ ,
\end{equation}
and then take the limit $r_{c}\rightarrow\infty$ in the regularized action.
Since in order to fulfill the Weak Energy Condition $\Lambda$ must be
negative, we can define
\begin{equation}
l_{0}:=\sqrt{-\frac{3}{\Lambda}}\ .\label{lcero}%
\end{equation}
The divergences in the regularized action cancel and further defining
$x_{+}:=\sqrt{\frac{\alpha}{\eta}}r_{+}$, the regularized action reduces to:
\begin{align}
I_{reg} &  =\frac{8\pi^{2}\kappa}{9}\frac{l^{2}x_{+}}{l_{0}^{2}\left(
2l_{0}^{2}+\left(  l^{2}+l_{0}^{2}\right)  x_{+}^{2}\right)  }\left[  3\left(
l^{2}-l_{0}^{2}\right)  ^{2}\arctan\left(  x_{+}\right)  \right.  \nonumber\\
&  \left.  +\left(  l^{2}-2l_{0}^{2}\right)  \left(  l^{2}+l_{0}^{2}\right)
x_{+}^{3}+3\left(  l_{0}^{4}-l^{4}+2l^{2}l_{0}^{2}\right)  x_{+}\right]  \ .
\end{align}
In terms of $l_{0}$ the W.E.C. imposes the following constraint:%
\begin{equation}
l^{2}>l_{0}^{2}\ ,
\end{equation}
this is, the AdS length of the asymptotic region ($l$), has to be larger than
the AdS length defined by the cosmological term in the action ($l_{0}$) given
by (\ref{lcero}).

As follow from the convention we are using defined in \cite{HP}, in the
canonical ensemble we relate the Euclidean action to the free energy by
$I_{reg}=\beta F$, therefore:%
\begin{equation}
M=\frac{\partial I_{reg}}{\partial\beta}\text{ and }S=\beta\frac{\partial
I_{reg}}{\partial\beta}-I_{reg}\ .
\end{equation}
Using our definition of the AdS radius $l$ as well as the definition of
$l_{0}$ given in (\ref{lcero}), the temperature is given by:%
\begin{equation}
T=\frac{\sqrt{3}x_{+}}{4\pi l}+\frac{\sqrt{3}l_{0}^{2}}{2\pi l\left(
l_{0}^{2}+l^{2}\right)  x_{+}}\ ,\label{temperature}%
\end{equation}
while the mass reads%
\begin{align*}
M &  =\frac{2}{3^{3/2}}\frac{\kappa\pi l}{l_{0}^{2}\left(  1+x_{+}^{2}\right)
\left(  l^{2}+l_{0}^{2}\right)  \left(  \left(  l^{2}+l_{0}^{2}\right)
x_{+}^{2}-2l_{0}^{2}\right)  }\left[  3\left(  1+x_{+}^{2}\right)  \left(
l^{2}-l_{0}^{2}\right)  ^{2}\left(  \left(  l^{2}+l_{0}^{2}\right)  x_{+}%
^{2}-2l_{0}^{2}\right)  \arctan(x_{+})\right.  \\
&  \left.  -2\left(  l^{2}-2l_{0}^{2}\right)  \left(  l^{2}+l_{0}^{2}\right)
^{2}x_{+}^{7}-2\left(  l^{2}+5l_{0}^{2}\right)  \left(  l^{2}-2l_{0}%
^{2}\right)  \left(  l^{2}+l_{0}^{2}\right)  x_{+}^{5}\right.  \\
&  \left.  +(l_{0}^{6}+7l^{4}l_{0}^{2}-13l^{2}l_{0}^{4}-3l^{6})x_{+}%
^{3}+6l_{0}^{2}(l^{2}-3l_{0}^{2})(l^{2}+l_{0}^{2})x_{+}\right]
\end{align*}
and the entropy reduces to:%
\begin{equation}
S=\frac{8\pi^{2}l^{2}\kappa x_{+}^{2}}{3l_{0}^{2}}\left[  \frac{\left(
l^{2}+l_{0}^{2}\right)  \left(  l^{2}-2l_{0}^{2}\right)  x_{+}^{4}+l_{0}%
^{2}\left(  l^{2}-l_{0}^{2}\right)  x_{+}^{2}+2l_{0}^{4}}{\left(  1+x_{+}%
^{2}\right)  \left(  2l_{0}^{2}-\left(  l_{0}^{2}+l^{2}\right)  x_{+}%
^{2}\right)  }\right]  \label{entropy}%
\end{equation}
One can further check that with these expressions the first law of black hole
thermodynamics%
\begin{equation}
dM=TdS\ ,
\end{equation}
is fulfilled.

For any positive $x_{+}$, there is a horizon and therefore from equation
(\ref{entropy}) one sees that requiring $S$ to be positive might induce some
restrictions on the couplings and $x_{+}$. Let us define the constant%
\begin{equation}
\xi:=\frac{l^{2}}{l_{0}^{2}}-1\ ,
\end{equation}
which must be strictly positive if we want to have a non-trivial real scalar
field outside of the horizon. The region in the plane $\xi$ vs $x_{+}$ in
which the entropy is positive is depicted in Figure 1. We see that for $\xi>1$
there is an upper bound on the radii of the black holes with positive entropy.
For $\xi<1$, there is a gap on the possible radii of the black holes with
positive entropy

\begin{figure}[h]
\centering
\includegraphics[scale=.7]
{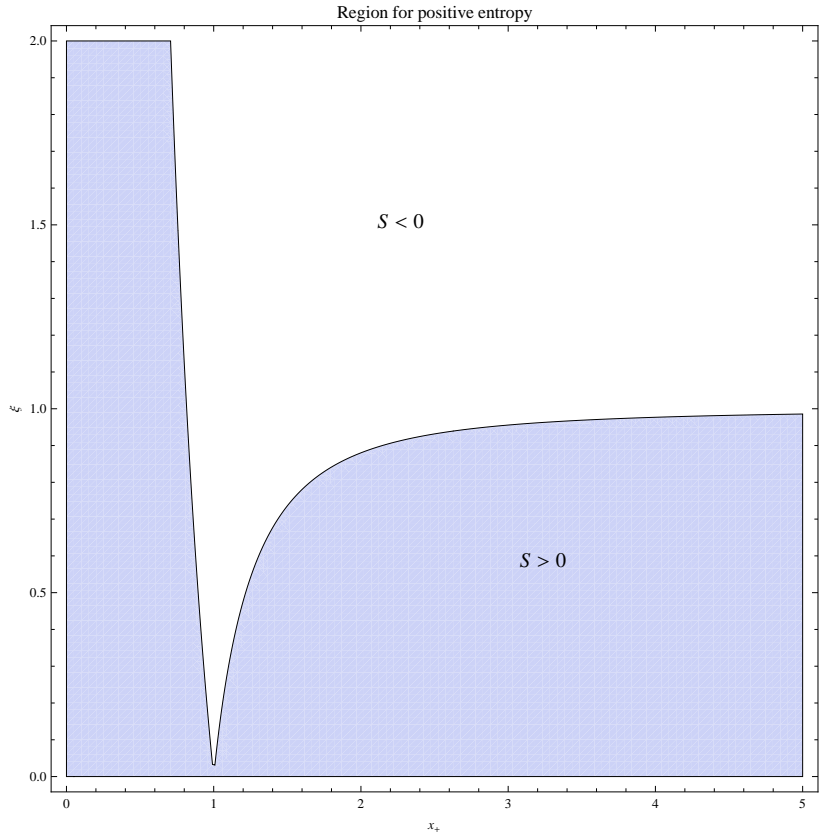}\caption{The grey region corresponds to the region with positive
entropy in the plane $\xi:=\frac{l^{2}}{l_{0}^{2}}-1$ vs $x_{+}$, while the
white regions stands for negative entropy. For $\xi>1$, requiring $S>0$
implies an upper bound on the black holes radii while for $0<\xi<1$ there is a
gap on the possible radii of black holes with positive entropy.}%
\label{fig1}%
\end{figure}

As it occurs for Schwarzschild-AdS in vacuum, the expression for the
temperature given in (\ref{temperature}) has a minimum at%
\begin{equation}
x_{+}=x_{0}:=\frac{2l_{0}}{\sqrt{2l_{0}^{2}+l^{2}}}\ ,
\end{equation}
where the temperature takes its minimum value%
\begin{equation}
T_{0}:=\frac{\sqrt{3}l_{0}}{\pi l\sqrt{2l_{0}^{2}+2l^{2}}}\ .
\end{equation}
Therefore there are no black holes with $T<T_{0}$, while for a given
temperature $T>T_{0}$ there are two possible black holes.

\bigskip

In the canonical ensemble the most probable configuration is the one with the
lowest Helmholtz free energy $F$. Figure [2] shows the four possible behaviors
of the free energy in terms of the temperature divided by the minimum
temperature ($T/T_{0}$) for the large (continuous line) and small (dashed
line) black holes.

\begin{figure}[h]
\centering
\includegraphics[scale=.7]
{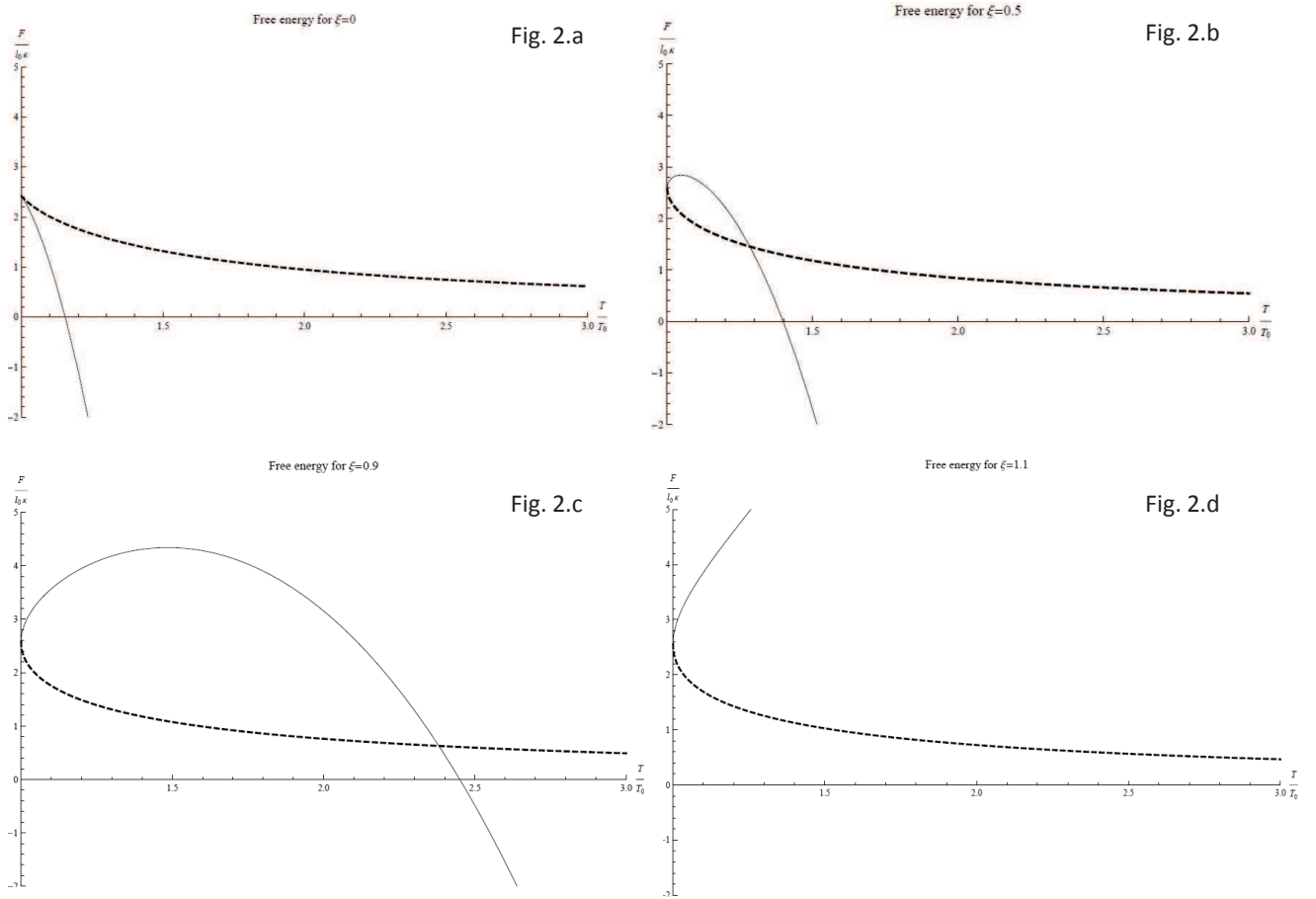}\caption{The free energy for large (continuous) and small (dashed)
black holes in terms of $\frac{T}{T_{0}}$, for $\chi=0$ (2.a), $\chi=0.5$ (2.b),
$\chi=0.9$ (2.c) and $\chi=1.1$ (2.d).}%
\label{fig2}%
\end{figure}

For $\xi=0$, the metric reduces to Schwarzschild-AdS therefore the situation
is like the one described by Hawking and Page in \cite{HP}. Small black holes
always have positive free energy, while for large black holes there is a
critical temperature above which the free energy becomes negative and the
thermal AdS (the thermal soliton with $\xi=0$) is less probable than the black
hole. For $0<\xi<1$ Figure 2.b and Figure 2.c show that small black holes
always have positive free energy and therefore they are less probable than the
thermal soliton, nevertheless there is a range of temperatures for which the
small black holes are more probable than large black holes. Above such
temperature (which fulfills a transcendental equation in terms of $\xi$) there
is a range in which the situation is like in G.R. in vacuum and therefore
large black holes suffer a phase transition at some critical temperature above
which the thermal soliton would tend to tunnel to black hole configurations.
For $\xi\geq1$, both the small and large black holes, have positive free
energy therefore both are less probable than the corresponding thermal
soliton, nevertheless large black holes are less probable than small black
holes, again in opposition to what occurs in G.R. in vacuum.

\bigskip

\section{Extending the solution to arbitrary dimensions $D$}

If one considers the theory defined in the action principle (\ref{action}) in
arbitrary dimensions, one can see that the metric%
\begin{equation}
ds^{2}=-F(r)dt^{2}+G(r)dr^{2}+r^{2}d\Sigma_{K,D-2}^{2}\ ,
\end{equation}
\ defines a solution with the following metric functions:%
\begin{align}
F_{D}\left(  r\right)   &  =-\frac{\mu}{r^{D-3}}+\left[  \left(  D-2\right)
K\eta-\frac{2\eta\Lambda}{\left(  D-1\right)  }r^{2}+\right.  \nonumber\\
&  \left.  \frac{\left(  \alpha+\eta\Lambda\right)  ^{2}}{\left(  D-3\right)
\left(  D-2\right)  \left(  D+1\right)  K\eta}r^{4}\ {}_{2}F_{1}\left(
1,\frac{D+1}{2},\frac{D+3}{2},-\frac{2\alpha}{\left(  D-3\right)  \left(
D-2\right)  K\eta}r^{2}\right)  \right]  \ ,\label{FD}\\
G_{D}\left(  r\right)   &  =\frac{A\left(  r\right)  }{B\left(  r\right)
}\frac{1}{F_{D}\left(  r\right)  }\ ,\\
\psi_{D}^{2}(r) &  =-\frac{4\kappa\left(  \Lambda\eta+\alpha\right)  r^{2}%
}{\eta\left(  2\alpha r^{2}+\eta\left(  D-2\right)  \left(  D-3\right)
K\right)  }G_{D}\left(  r\right)  \ .\nonumber
\end{align}
Here we have defined%
\begin{align*}
A\left(  r\right)   &  :=\left[  K^{2}\eta^{3}\left(  D+3\right)  \left(
D-3\right)  ^{2}\left(  D-2\right)  ^{2}\left(  \left(  D-2\right)  \left(
D-3\right)  K-2r^{2}\Lambda\right)  +\right.  \\
&  \left.  r^{2}\left(  \alpha+\eta\Lambda\right)  ^{2}\left(  D-2\right)
\left(  K\eta\left(  D-2\right)  \left(  D-3\right)  \left(  D+3\right)
\ {}_{2}F_{1}\left(  1,\frac{D+1}{2},\frac{D+3}{2},-\frac{2\alpha r^{2}%
}{\left(  D-3\right)  \left(  D-2\right)  K\eta}\right)  \right.  \right.  \\
&  \left.  \left.  -4\alpha\ {}_{2}F_{1}\left(  2,\frac{D+3}{2},\frac{D+5}%
{2},-\frac{2\alpha r^{2}}{\left(  D-3\right)  \left(  D-2\right)  K\eta
}\right)  \right)  \right]  \\
B\left(  r\right)   &  :=\left(  D+3\right)  \eta K^{2}\left(  D-2\right)
\left(  D-3\right)  ^{2}\left(  \left(  D-3\right)  \left(  D-2\right)
K\eta+2\alpha r^{2}\right)  \ .
\end{align*}
The constant $\mu$ is the only integration constant of the solution
and\ $_{2}F_{1}$ is the hypergeometric function. In order to obtain this
solution we had set to zero the integration constant appearing in the first
integral of the equation of the field as in four dimensions. Let us define the
dimensionless coordinate $\rho$ by%
\begin{equation}
\rho:=\sqrt{\frac{2\alpha}{\left(  D-3\right)  \left(  D-2\right)  K\eta}}r\ .
\end{equation}
The hypergeometric function appearing in equation (\ref{FD}) can be rewritten
in terms of elementary functions in the following manner. For even dimensions
$D=2n$, we have%
\begin{equation}
_{2}F_{1}\left(  1,n+\frac{1}{2},n+\frac{3}{2},-\rho^{2}\right)  =\left(
-1\right)  ^{n}\left(  2n+1\right)  \rho^{-2n-2}\left[  \rho\arctan\rho
+\sum\limits_{j=1}^{n}\frac{\left(  -1\right)  ^{j}}{2j-1}\rho^{2j}\right]
\ ,
\end{equation}
while for odd dimensions $D=2n+1$, we use%
\begin{equation}
_{2}F_{1}\left(  1,n+1,n+2,-\rho^{2}\right)  =\left(  -1\right)  ^{n}\left(
n+1\right)  \rho^{-2n-2}\left[  \ln\left(  \rho^{2}+1\right)  +\sum
\limits_{j=1}^{n}\frac{\left(  -1\right)  ^{j}}{j}\rho^{2j}\right]  \ .
\end{equation}

As in four dimensions, it can be seen that these metrics describe
asymptotically locally AdS black holes with real scalar fields in the domain
of outer communication.

\section{Asymptotically flat black holes supported by the Einstein-kinetic
coupling}

Here we will show that the inclusion of a cosmological term in the action
allows finding a new asymptotically locally flat black hole. In this case we
will consider that the matter sector is given only by the kinetic term of the
scalar field which is constructed with the Einstein tensor, namely we shall
consider the action (\ref{action}) with $\alpha=0$. Thus, the action reduces
to%
\begin{equation}
I[g_{\mu\nu},\phi]=\int\sqrt{-g}d^{4}x\ \left[  \kappa\left(  R-2\Lambda
\right)  +\frac{\eta}{2}G_{\mu\nu}\nabla^{\mu}\phi\nabla^{\nu}\phi\right]  \ .
\end{equation}
In the same manner than before, the equation for the field allows a first
integral which brings in an integration constant. If we set to zero such
integration constant, then we find that for $K\neq0$, the following metric
defines a solution of the system%
\begin{equation}
ds^{2}:=-H\left(  r\right)  dt^{2}+\frac{15\left(  \Lambda r^{2}-2K\right)
^{2}}{K}\frac{dr^{2}}{H\left(  r\right)  }+r^{2}d\Sigma_{K,2}^{2}\ ,
\end{equation}
provided%
\begin{equation}
H\left(  r\right)  :=\left(  60K^{2}-20\Lambda Kr^{2}+3\Lambda^{2}%
r^{4}\right)  -\frac{\mu}{r}\text{\ ,}%
\end{equation}
and the derivative of the scalar field is given by%
\begin{equation}
\psi_{0}^{2}(r):=-\frac{30\kappa\Lambda r^{2}\left(  \Lambda r^{2}-2K\right)
^{2}}{\eta K^{2}H\left(  r\right)  }\ .
\end{equation}
\bigskip The following comments are in order:

- As in the previous cases, at a possible non-degenerate horizon $r=r_{+}$ the
lapse function $H\left(  r_{+}\right)  $ vanishes, and therefore the scalar
field vanishes but it's non-analytic at that point.

- For vanishing $\Lambda$, the scalar field it self vanishes and the solution
reduces to the topological Schwarzschild solution in flat space, which
represents a black hole only in the spherically symmetric case ($K=1$).

- This solution is asymptotically locally flat since%
\begin{equation}
R_{\ \ \lambda\rho}^{\mu\nu}\underset{r\rightarrow\infty}{\sim}0\ .
\end{equation}

- In order to have a real scalar field in the region where $H\left(  r\right)
$ is positive, i.e. outside a possible event horizon, we need to impose
$\Lambda/\eta<0$.

- For non-vanishing $\mu$, the metric has a singularity at the origin.

- When $K$ and $\Lambda$ have the same sign, there is a curvature singularity
at a finite radius $r=r_{s}:=\sqrt{2K/\Lambda}$, for any value of $\mu$.

- For $\mu=0$, the point $r=0$ is a symmetric center and therefore the range
of the radial coordinate is $r\geq0$. If even more in this case $K/\Lambda<0$
the metric is regular everywhere and therefore describes an asymptotically
locally flat gravitational soliton.

- When $K/\Lambda<0$ the singularity at $r=r_{s}$ disappears and the
singularity at the origin is surrounded by an event horizon provided $\mu>0$.
In this case the solution represents an asymptotically locally flat black hole.

- For $K/\Lambda>0$ the singularity at $r=r_{s}>0$ is surrounded by an event
horizon provided $\mu>\mu_{c}$ for certain critical value of $\mu$. In this
case as well, the spacetime represents an asymptotically locally flat black hole.

\bigskip

The case $K=0$ (and $\alpha=0$) integrates in a different manner. As before,
in order to have a vanishing integration constant from the first integral of
the equation of the field, in addition to the field equations we need to
impose $G^{rr}=0$. The field equations therefore imply that $\Lambda$ has to
vanish and the metric and the derivative of the scalar field ($\psi_{0}\left(
r\right)  =\phi^{\prime}\left(  r\right)  $) are given by%
\begin{align}
ds^{2} &  =-\frac{C}{r}dt^{2}+\frac{dr^{2}}{G\left(  r\right)  }+r^{2}\left(
dx^{2}+dy^{2}\right)  \ ,\\
\psi\left(  r\right)  ^{2} &  =-\frac{4\kappa}{\eta}G\left(  r\right)
+J\frac{G\left(  r\right)  ^{3/2}}{\sqrt{r}}\ ,
\end{align}
where $C$ and $J$ are integration constants and $G\left(  r\right)  $ is an
arbitrary function. This is a degenerate case in which the system is
under-determined since one of the metric functions is completely arbitrary.
Note that the function $G\left(  r\right)  $ cannot be absorbed by a diffeomorphism.

\section{Discussion}

We have found a new family of asymptotically AdS and locally flat black holes
supported by scalar field in arbitrary dimensions. The theory we considered is
a particular case of Horndeski theory and therefore the field equations and
the energy-momentum tensor are of second order. For the ansatz considered, the
equation for the field allows for a first integration giving rise to an
integration constant. Following the steps of Rinaldi's work \cite{Rinaldi} we
were able to obtain an exact solution imposing such integration constant to
vanish. This imposes an extra constraint in the geometry that turns out to
open a new family of non-trivial solutions. The inclusion of a cosmological
term in the action allowed us to find a solution with a real scalar field
outside the horizon, and allowed us finding asymptotically locally flat black
hole solutions as well. It's important to note that the cosmological constant
at infinity is not given by the cosmological $\Lambda$ term in the action but
in terms of the couplings $\alpha$ and $\eta$ that appear in the kinetic term
of the field (see \ref{defl}).

\bigskip

The solutions are not continuously connected with the maximally symmetric AdS
or flat backgrounds since the scalar field cannot be turned off by setting the
single integration constant to some value. Nevertheless, since our family of
metrics contains a further integration constant, it is possible to show that
within such a family there is a unique regular spacetime. Such spacetime is a
gravitational soliton and we use it in the four dimensional, spherically
symmetric case to defined a regularized Euclidean action and explore the
thermodynamics of the black hole solution. A similar situation occurs with the
AdS soliton, which can be considered as the background for some topological
AdS black holes, as well as in gravity in 2+1 with scalar fields, where the
gravitational solitons are the right backgrounds to consider to give a
microscopic description of the black hole entropies \cite{Correa1}%
-\cite{Correa3}.

\bigskip

The thermodynamics of the asymptotically AdS black holes depends strongly on
the ratio between the AdS length of the asymptotic region $l$ which is
determined by the couplings of the scalar and the "bared" AdS length $l_{0}$
constructed in terms of the cosmological term in the action $\Lambda$. We have
shown that in a certain region of the space of parameters, the thermal soliton
will tend to tunnel to large black holes. In opposition to what occurs in
General Relativity, there is also a range of temperatures for which small
black holes are more probable than large black holes.

\bigskip

Even though it is probably not possible to give an exact, closed form for the
general static black hole solution of Horndeski theory, there are some
particular subsets of theories that might offer the possibility of finding
exact solutions. For example, in dimensions higher than four, since the
Lovelock tensors are non-trivial, it would be interesting look for black holes
with scalar field in which the kinetic couplings are constructed with such
Lovelock tensors. In some of those theories, the kinetic term will be
constructed with the field equations of a gravity theory that has an extra
symmetry \cite{Zreview}, and this might help in the integration of the field
equations. Other theories that belong to Horndeski family are theories with
invariance under local Weyl rescaling. In reference \cite{ORconf} it was shown
that it is possible to construct theories that generalize the usual
conformally coupled scalar field, including couplings of the scalar with Euler
densities of higher degree (e.g. $\phi^{4}\left(  R^{2}-4R_{\mu\nu}R^{\mu\nu
}+R_{\alpha\beta\gamma\delta}R^{\alpha\beta\gamma\delta}\right)  $). This was
done by introducing a four-rank tensor which contains the curvature and
derivatives of the field and transforms covariantly under local Weyl
rescaling\footnote{In four dimensions a different approach for constructing
theories with non-minimal coupling has been recently given in \cite{Minas},
allowing two scalar fields as well.}. In \cite{ORconf} it was shown that
restricting to Lovelock-like combinations in such four-rank tensor ensures the
field equations and energy-momentum tensor to be of second order and therefore
those combinations are a subclass of Horndeski's theories, which in this case
have an extra local symmetry that might be useful in the integration of the
field equations.

\bigskip

\section{Acknowledgments}

We thank Fabrizio Canfora, Francisco Correa, Alex Giacomini, Gaston Giribet,
Nicolas Grandi, Mokhtar Hassaine and Sourya Ray for many useful conversations.
We would like to thanks as well the referee of this manuscript for a thorough
revision and usefull coments. Research of A.A. is supported in part by the
FONDECYT grant 11121187 and by the CNRS project \textquotedblleft Solutions
exactes en pr\'{e}sence de champ scalaire\textquotedblright. A.C. thanks the
support of Beca CONICYT de Doctorado Nacional 2010. J. O. is supported in part
by FONDECYT grant 1141073. A.C. and J.O. appreciate the hospitality of the
International Center of Theoretical Physics (ICTP) where part of this work was done.

\end{document}